\documentclass[fleqn,usenatbib]{mnras}

\usepackage{newtxtext,newtxmath}

\usepackage[T1]{fontenc}

\DeclareRobustCommand{\VAN}[3]{#2}
\let\VANthebibliography\thebibliography
\def\thebibliography{\DeclareRobustCommand{\VAN}[3]{##3}\VANthebibliography}

\usepackage{graphicx}	
\usepackage{amsmath}	
\usepackage{amssymb}	

\newcommand{\hii}{H~{\sc ii}}	

\title[Recombination in PNe]{The impact of strong recombination on temperature determination in planetary nebulae}

\author[V. Gomez-Llanos et al.]{
V. Gomez-Llanos,$^{1}$\thanks{E-mail: vgomez@astro.unam.mx}
C. Morisset,$^{1}$
J. Garc\'ia-Rojas,$^{2,3}$
D. Jones,$^{2,3}$
R. Wesson,$^{4}$
\newauthor
R. L. M. Corradi,$^{5,2}$
and H. M.J. Boffin$^{6}$
\\
$^{1}$Instituto de Astronomia (IA), Universidad Nacional Aut\'onoma de M\'exico,
Apdo. postal 106, C.P. 22800 Ensenada, 
Baja California, M\'exico.\\
$^{2}$Instituto de Astrof\'isica de Canarias, E-38200, La Laguna, Tenerife, Spain\\
$^{3}$Universidad de La Laguna. Depart. de Astrof\'isica, E-38206, La Laguna, Tenerife, Spain\\
$^{4}$Department of Physics and Astronomy, University College London, Gower St, London WC1E 6BT, UK\\
$^{5}$GRANTECAN, Cuesta de San Jos\'e s/n, E-38712, Bre\~na Baja,  La Palma, Spain  \\
$^{6}$European Southern Observatory, Karl-Schwarzschild-Str. 2, 85748 Garching bei Muenchen, Germany 
}

\date{Accepted XXX. Received YYY; in original form ZZZ}

\pubyear{2015}

\begin{document}
\label{firstpage}
\pagerange{\pageref{firstpage}--\pageref{lastpage}}
\maketitle

\begin{abstract}
The long-standing difference in chemical abundances determined from optical recombination lines and collisionally excited lines raises questions about our understanding of atomic physics, as well as the assumptions made when determining physical conditions and chemical abundances in astrophysical nebulae. Here, we study the recombination contribution of [O~{\sc iii}]~4363 and the validity of the line ratio [O~{\sc iii}]~4363/4959 as a temperature diagnostic in planetary nebulae with a high abundance discrepancy. We derive a fit for the recombination coefficient of [O~{\sc iii}]~4363 that takes into account the radiative and dielectronic recombinations, for electron temperatures from 200 to 30,000~K. We estimate the recombination contribution of [O~{\sc iii}]~4363 for the planetary nebulae Abell~46 and NGC~6778 by subtracting the collisional contribution from the total observed flux. We find that the spatial distribution for the estimated recombination contribution in [O~{\sc iii}]~4363 follows that of the O~{\sc ii}~4649 recombination line, both peaking in the central regions of the nebula, especially in the case of Abell~46 which has a much higher abundance discrepancy. The estimated recombination contribution reaches up to 70\% and 40\% of the total [O~{\sc iii}]~4363  observed flux, for Abell~46 and NGC~6778, respectively.
\end{abstract}




\section{Introduction} 
\label{sec:intro}

When measuring chemical abundances from faint heavy-element optical recombination lines (ORLs), it is found that they are always greater than those measured from the much brighter collisionally excited lines (CELs). Being known for more than 70 years, this \emph{abundance discrepancy problem} is probably the most important challenge to our understanding of the physics of photoionized nebulae \citep[see][and references therein]{2019_Garcia-Rojas_arXiv}. Several scenarios have been proposed to resolve the issue, the two most popular being: i) the existence of temperature fluctuations within a chemically homogeneous plasma \citep{1967Peimbert_apj150,  1980Torres-Peimbert_apj238}, and ii) the presence of  cold, metal-rich gaseous clumps in the nebula, which are very efficiently cooled by the heavy elements \citep{2000Liu_mnra312}. 

However, neither of these scenarios seem appropriate to universally explain the complete range of abundance discrepancy factors (ADFs, i.~e. the ratio between the abundances determined from ORLs and CELs) observed in both {\hii}  regions and planetary nebulae \citep[PNe; see][]{2018Wesson_mnra480}\footnote{An updated database on the ADFs measured in {\hii} regions and PNe can be found at \url{https://www.nebulousresearch.org/adfs/}}. Moreover, the mechanisms producing, and allowing for the survival of, temperature fluctuations in a photoionized plasma are still under debate  \citep{2017Peimbert_pasp129}, while the same is true for the physical origin of the metal-rich component \citep{2007Stasinska_aap471, 2015Corradi_apj803}. Nevertheless, some observational evidence of the existence of two or more gaseous phases in PNe have been found by several authors in recent years \citep{2003Wesson_mnra340, 2006Liu_mnra368, 2008Wesson_mnra383, 2013Richer_apj773, 2017Richer_aj153, 2017Pena_mnras472}. 

In particular, PNe with ADFs $> 10$ have proven to be very interesting objects, as their extreme ADFs seem to be linked with the evolution of a central close-binary system that has experienced a common envelope phase \citep{2006Liu_mnra368, 2015Corradi_apj803, 2016Garcia-Rojas_apjl824,2016Jones_mnra455, 2018Wesson_mnra480}, even if the nature of the relationship is still a mystery. Detailed analysis of the physical conditions and chemical abundances of a few objects has led several authors to suggest that the ionized gas comprises two different phases: an H-rich phase, which is dominated by Hydrogen and Helium recombination lines and collisionally excited lines (CELs) from heavy elements (O, N, Ne, Ar...), alongside a much colder, H-poor phase with strong emission in the optical recombination lines (ORLs) of heavy elements (C, N, O, Ne) and almost no CEL emission \citep{2000Liu_mnra312, 2005Wesson_mnra362, 2015Corradi_apj803, 2018Wesson_mnra480}. Under this hypothesis, accurately determining the physical conditions (electron temperature, $T_e$, and electron density, $n_e$) from different CEL and ORL diagnostics is crucial to properly determine the chemical abundances in each phase. However, having two gas phase components with different chemical contents in an ionized gas complicates the computation of physical conditions and chemical abundances from an observational point of view. 

A first estimate of how the presence of multiple gas components could affect the determination of physical conditions and chemical abundances in the main nebular shell was made by \citet{2000Liu_mnra312}, who computed new recombination coefficients for the $T_e$-sensitive [N~{\sc ii}] $\lambda$5754 and [O~{\sc ii}] $\lambda\lambda$ 7320+30 auroral lines and found that recombination excitation was important in exciting these lines and that ignoring it would lead to an overestimated $T_e$. These authors also proposed a fit to the contribution of radiative recombination to the widely used [O~{\sc iii}] $\lambda$4363 auroral line, valid for $T_e > 8,000~K$.

In this letter we want to explore the classical [O~{\sc iii}] $\lambda$4363/$\lambda$4959 $T_e$ diagnostic, that can be strongly contaminated by recombination in extreme ADF PNe and, therefore, is no longer suitable for the measurement of $T_e$. In this work we try to determine the contribution of the recombination to the [O~{\sc iii}] $\lambda$4363 line, for the PNe NGC\,6778 and Abell\,46. In \S~\ref{sec:observations} we briefly describe the observational data used in this paper; in \S~\ref{sec:oiiiemissivity} we present new calculations to compute the recombination contribution to the [O~{\sc iii}] $\lambda$4363 CEL emissivity; in \S~\ref{sec:biabund} we estimate the recombination contribution from an observational point of view and, finally, in \S~\ref{sec:discussion} we discuss our results.

\section{Observations} \label{sec:observations}

We have used long-slit, intermediate-resolution spectra taken by our group of the extreme ADF planetary nebulae NGC\,6778 \citep[with FORS2-VLT 8.2m; see][ADF$\sim$18]{2016Jones_mnra455} and Abell\,46 \citep[with ISIS-WHT 4.2m; see][ADF$\sim$120]{2015Corradi_apj803} respectively. The FORS2 observations covered the wavelength range 3600-5000 \AA\   with an average spectral resolution of 1.5 \AA. The ISIS observations covered the wavelength range 3610-5050 \AA\   with a spectral resolution of 0.8 \AA. For additional details on the observations and data reduction, we refer the reader to the original references.

For each long slit, we split the 2D-spectrum into several spatial bins along the slit --2$''$.5 and 0$''$.5 wide for Abell\,46 and NGC\,6778, respectively-- which provides enough signal to noise for the faintest lines of interest to be measured. The fluxes of the [O~{\sc iii}] $\lambda$4363 and $\lambda$4959 CELs and the O~{\sc ii} $\lambda\lambda$4649+50 ORL are obtained by automatically fitting Gaussian profiles to each line (for the ORL line, a double Gaussian is used to take into account the two members of the multiplet at 4649.13~\AA\ and 4650.25~\AA; we ignore the contribution of C
~{\sc iii} $\lambda$4650.25 because other lines of the same multiplet such as C
~{\sc iii} $\lambda$4647.42 have not been reported in the literature spectra of either object \citep{2016Jones_mnra455,2015Corradi_apj803}. The uncertainties are determined in each spatial bin through a quadratic mean of the difference between the Gaussian fit and the signal. 

\begin{figure*}
\includegraphics[scale = 0.5]{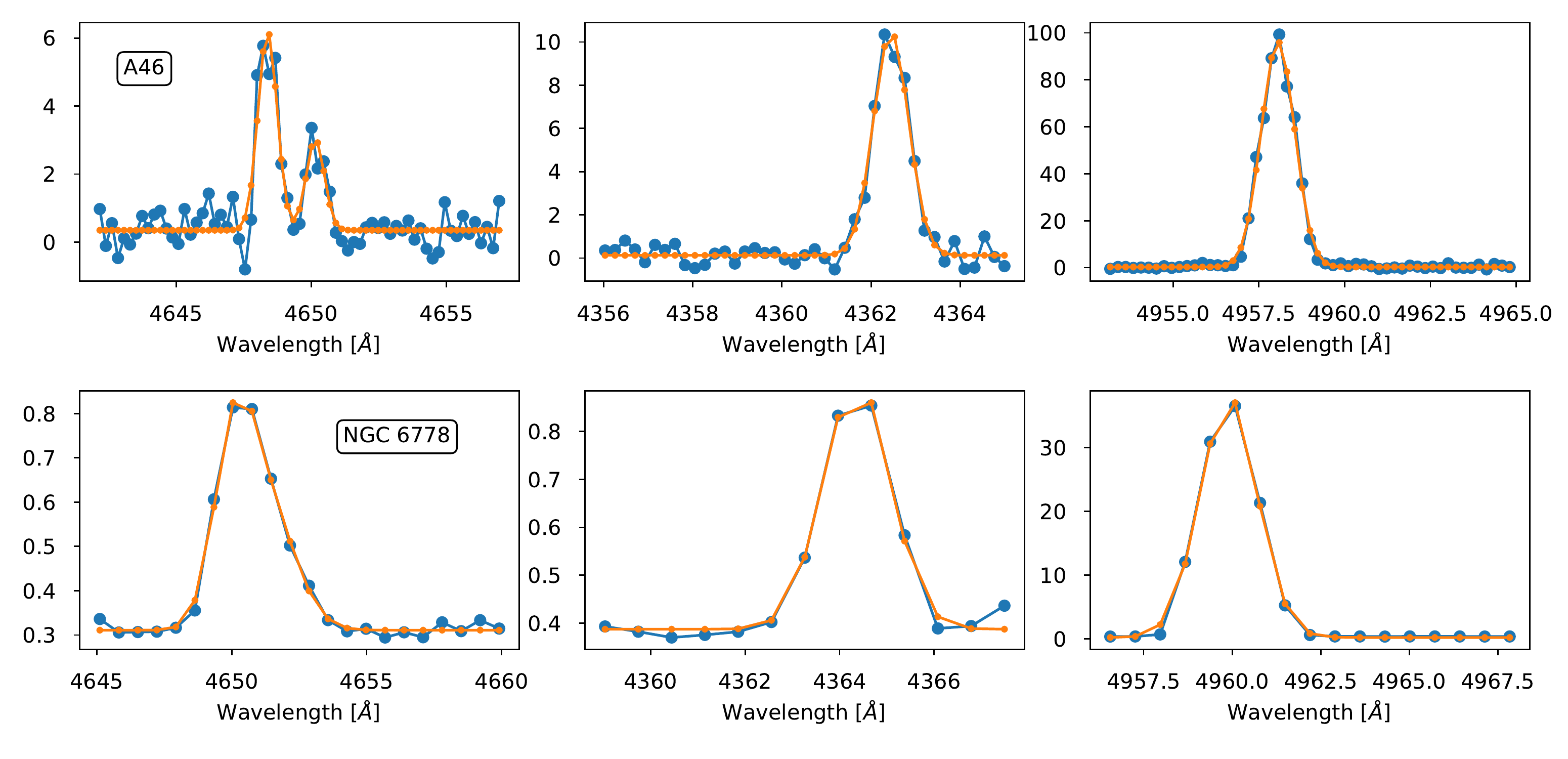}
\caption{Examples of the fit to the emission lines. Upper panels for A46 and lower panels for NGC 6778. From left to right, the fits are for the O~{\sc ii} $\lambda\lambda$4649+50, the [O~{\sc iii}] $\lambda$4363 and the [O~{\sc iii}] $\lambda$4959 emission lines. Line fluxes are in arbitrary units.\label{fig:linefit}}
\end{figure*}

In Fig.~\ref{fig:linefit}, we show examples of the line fitting process for the 2 PNe considered in this paper (upper panels for Abell\,46 and lower panels for NGC\,6778). 

\section{The limits in computing [O~{\sc iii}] $\lambda$4363 intensity}\label{sec:oiiiemissivity}

To compute the [O~{\sc iii}] $\lambda$4363 emission, we first consider the contribution from the radiative deexcitation of the O$^{++}$ ion following an excitation of the level $^1\mathrm{S}_0$ by collision with a free electron of the plasma. This is obtained using PyNeb \citep{2015Luridiana_aap573} version 1.1.10, based on collision strengths by \citet{2014Storey_mnras441} and transition probabilities by \citet{2004Froese-Fischer_Atom87}. 

We also have to carefully take into account the radiative recombination (RR) computed by the fit from \citet{1991Pequignot_aap251} and the dielectronic recombination (DR) from \citet{1984Nussbaumer_aaps56}. 

A fit for the total recombination contribution is given by \citet{2000Liu_mnra312} as: 
\begin{equation} \label{eq:liu}
\frac{I(4363)}{I(H\beta)} = 12.4 \times t^{0.59} \times \frac{O^{3+}}{H^+}.
\end{equation} 
This line ratio [O~{\sc iii}] $\lambda$4363/H$\beta$ leads to a recombination coefficient of the single [O~{\sc iii}] $\lambda$4363 line close to:

\begin{equation} 
\alpha_{4363} [\mathrm{cm^3 s^{-1}}] \simeq 3.3 \times 10^{-13} \times t^{-0.21}, 
\end{equation}

where t = T$_e$/$10^4$~K and considering $\alpha_{{\rm H}\beta} \simeq 2.94\times 10^{-14} \times t^{-0.80}\  \mathrm{cm^3 s^{-1}}.$ 

This fit has been obtained when the DR has a considerable effect on the total recombination. This occurs  for $T_e$ between 8,000~K and 20,000~K. For $T_e <$  5,000~K, the DR is negligible compared to the RR and the dependency on $T_e$ does not follow the fit by \citet{2000Liu_mnra312} anymore. We computed a new fit that reproduces the sum RR+DR within 3\% from 200 to 30,000~K:

\begin{equation} \label{eq:fit}
\alpha_{4363} \simeq 2.63\times 10^{-13} \times t^{-0.6} + 1.4\times 10^{-13} \times e^{-0.8/t}.
\end{equation} 

\begin{figure}
\includegraphics[scale = 0.55]{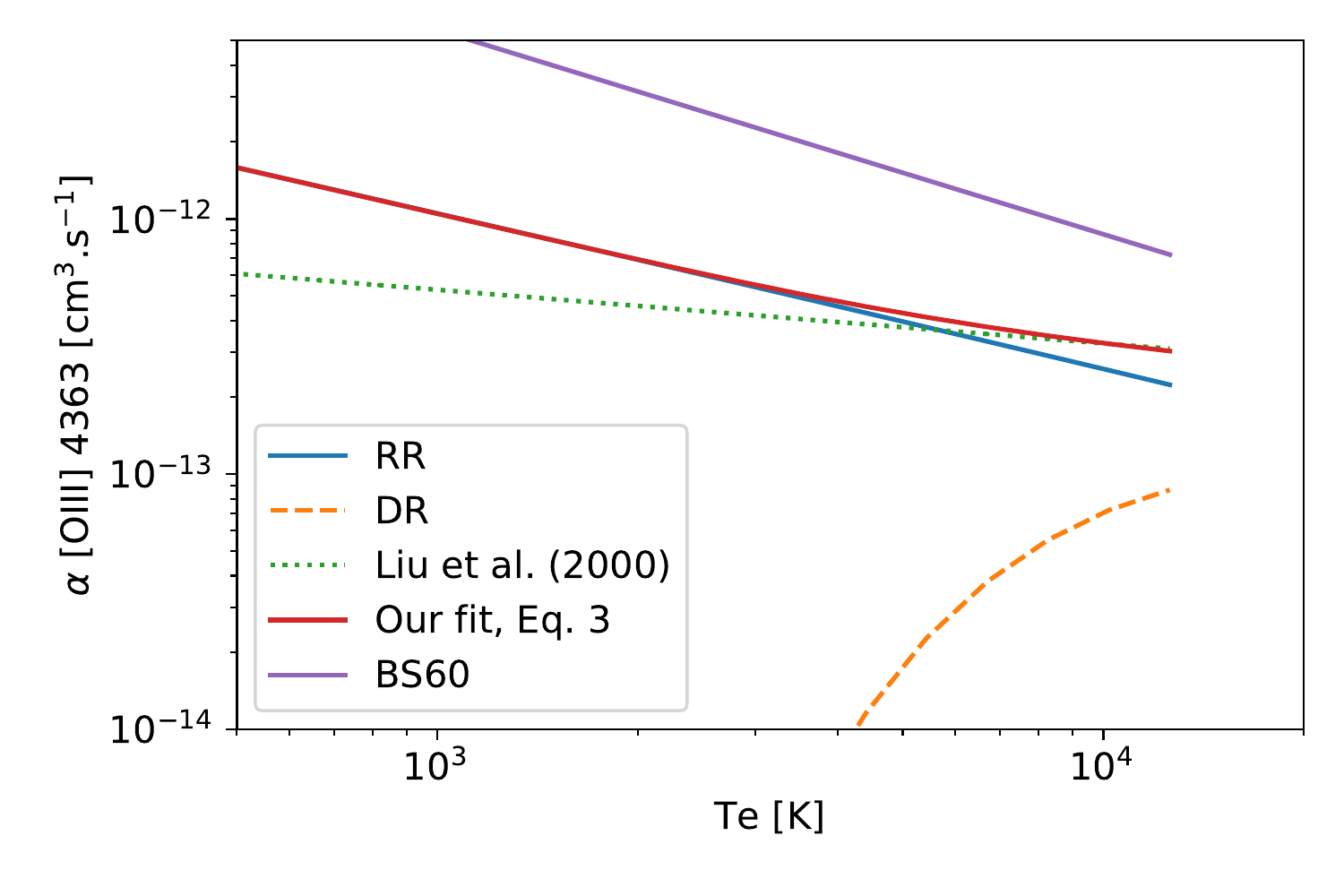}
\caption{Recombination coefficients of [O~{\sc iii}] $\lambda$4363: radiative recombination (RR) computed by \citet{1991Pequignot_aap251} (blue line), dielectronic recombination (DR) by \citet{1984Nussbaumer_aaps56} (orange dashed line), the value obtained by the formula from \citet{2000Liu_mnra312} (Eq.~\ref{eq:liu}, green dot line) and our fit to RR+DR (Eq.~\ref{eq:fit}, red dot-dashed line). The actual value of RR+DR is not shown, as it is not distinguishable from our fit. The BS60 values from \citet[][]{1960_Burgess_mnra120} upper limits used in Cloudy are also shown in purple.\label{fig:alphas}}
\end{figure}

This fit is used in the multipurpose photoionization code Cloudy \citep{2017Ferland_rmxaa53} since v17.02. 
In Fig.~\ref{fig:alphas}, we can see the variation of recombination coefficients with $T_e$ for the RR and DR computed from \citet{1991Pequignot_aap251} and \citet{1984Nussbaumer_aaps56} respectively, as well as the fit by \citet{2000Liu_mnra312} and our fit from Eq.~\ref{eq:fit}.
Note that the recombination computed in Cloudy until v17.02 is rather overestimated as it uses \citet{1960_Burgess_mnra120} upper limits (purple line in Fig.~\ref{fig:alphas}). 

It is important to notice here that in general, no simple fit to the line ratio [O~{\sc iii}] $\lambda$4363/H$\beta$ (like the one obtained by \citet{2000Liu_mnra312}, Eq.~\ref{eq:liu} above) can be obtained, as the regions where the recombination lines [O~{\sc iii}] $\lambda$4363 and H$\beta$ are produced can be very different, especially in terms of temperature, densities and volume. 
The complete relation is :
\begin{equation}\label{eq:I4363b_int}
   \frac{I(4363)}{I(\beta)} = \frac{\int_V E_{4363}\ \alpha_{4363}(T)\ n(O^{3+})\ n(e)\ dV} 
                              {\int_V E_{\beta}\ \alpha_{\beta}(T)\ n(H^{+})\ n(e)\ dV},
\end{equation}
where $E_\lambda$ is the energy of the corresponding emission line and $n(X)$ is the density (by number) of the ion $X$ responsible for the line emission (namely $O^{3+}$ and $H^+$ in the present case). 

\section{Testing bi-metallicity hypothesis} \label{sec:biabund}

In the following we explore the case where the nebula is made of two regions of very different abundances. Then, the H$\beta$ line is mainly emitted by the close-to-solar metallicity region 1 and the [O~{\sc iii}] $\lambda$4363 recombination line is mainly emitted by a cold, metal-rich region 2,  equation~\ref{eq:I4363b_int} leads to:

\begin{equation}\label{eq:I4363b_12}
    \frac{I(4363)}{I(\beta)} = \frac{E_{4363}\ \alpha_{4363}(T_2)\ n(O^{3+})_2\ n(e)_2\ V_2} 
                               {E_{\beta}\ \alpha_{\beta}(T_1)\ n(H^{+})_1\ n(e)_1\ V_1}, 
\end{equation}

where the 1 and 2 subscripts indicate a mean value over the regions 1 and 2 respectively\footnote{In the general situation, both lines are emitted by both regions and the line intensities are obtained by summing contributions from region 1 and 2, leading to an even more complex equation for the line ratio: 
\begin{equation}\label{eq:I4363b_general}
        \frac{I(4363)}{I(\beta)} = \frac{E_{4363}\ \left[ I(4363)_1 + I(4363)_2\right]}
                                    {E_{\beta}\ \left[ I(\beta)_1 + I(\beta)_2\right]},
\end{equation}
where $I(\lambda)_i = \alpha_\lambda(T_i)\ n(X)_i\ n(e)_i\ V_i$.
}.

Therefore, if $T_1$ and $T_2$ are very different, the simplification of the temperature dependent power terms in the recombination coefficients of the two lines cannot be applied, nor do the $n_e$ ratio $n(e)_1/n(e)_2$ and the volume ratio $V_1/V_2$ cancel. These ratios cancel only if the same region of the nebula is considered to emit both lines. The abundance ratio $O^{3+}/H^+$ only appears in a final relation if the implicit hydrogen densities $n(H)_1$ and $n(H)_2$ are the same. 

\citet{2020Gomez-Llanos} explored a case where an ADF(O$^{2+}$) $\sim$ 8, determined from observations of NGC 6153, can be reproduced by models in which the actual abundance ratio between the two components (termed the abundance contrast factor or ACF) is as high as 600. In their Annex, they even show that an ACF of 1,000 could lead to an apparent ADF of 1!

As derived from Eqs.~\ref{eq:I4363b_12} and \ref{eq:I4363b_general} above, it is very difficult to determine the contribution to the emission of the [O~{\sc iii}] $\lambda$4363 that comes from the recombination in cases where the gas has two phases of different metallicities, with the metal recombination contribution mainly coming from the H-poor region. Estimating the parameters (e.g. $T_e$ and $n_e$) of both regions needed in Eq.~\ref{eq:I4363b_12} is very hard, as for most objects the observed morphology does not allow to separate the emission coming from each region. 

We have seen that, from a theoretical stand-point, it is almost impossible to correctly determine the contribution to the emission of the [O~{\sc iii}] $\lambda$4363 line that originates from recombination. We can nevertheless attempt to obtain this contribution on an observational basis. In the following, we try to determine the recombination contribution by removing the contribution of the collisionally excited emission to the total emission.

Several authors have found that the spatial distribution of the [O~{\sc iii}] $\lambda$4363 emission is very similar to that of the O~{\sc ii} $\lambda$4649 one, but very different from the [O~{\sc iii}] $\lambda$5007, 4959 lines \citep{2015Corradi_apj803, 2016Garcia-Rojas_apjl824, 2016Jones_mnra455, 2018Wesson_mnra480}. The observed behavior is consistent with an increasing temperature towards the central parts of the PN, which is at odds with the fact that O~{\sc ii} ORL emission also peaks at the center of the nebula, indicating that the cold, H-poor gas is located close to the central star \citep{2016Garcia-Rojas_apjl824}.

\begin{figure*}
\centering
\includegraphics[scale = 0.28]{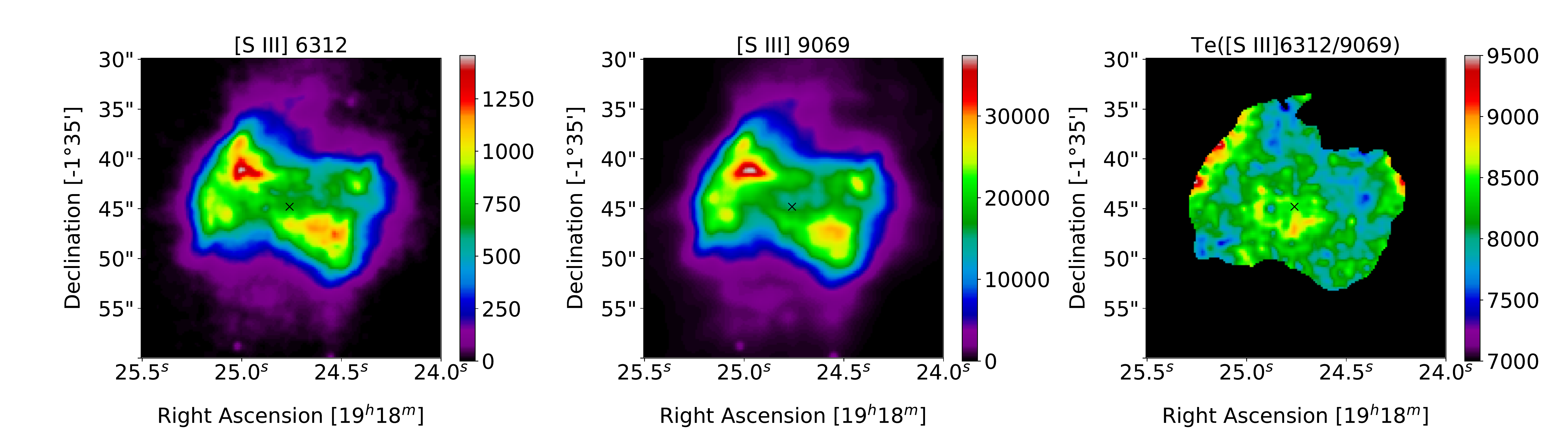}
\caption{Left and middle panels: MUSE emission line derredened maps of the $T_e$ sensitive [S~{\sc iii}] $\lambda$6312 and [S~{\sc iii}] $\lambda$9068 CELs for NGC 6778, showing a very similar spatial distribution. The raw maps have been convolved with a gaussian kernel with $\sigma$=1.5 pixels. The ``x'' marks the position of the central star in both maps. The colorbar shows the flux in units of 10$^{-20}$ erg s$^{-1}$ cm$^{-2}$ \AA$^{-1}$. Right panel: $T_e$([S~{\sc iii}]) map computed with \textit{PyNeb} from the extinction corrected [S~{\sc iii}] $\lambda$6312/$\lambda$9068 line ratio (a cut is made for intensities lower than 10\% of the peak emission in [S~{\sc iii}] $\lambda$6312). The average $T_e$ weighted by the flux of the $\lambda$6312 line is $\sim$8150 K. 
\label{fig:siii_lines}}
\end{figure*}

\begin{figure}
\includegraphics[scale = 0.45]{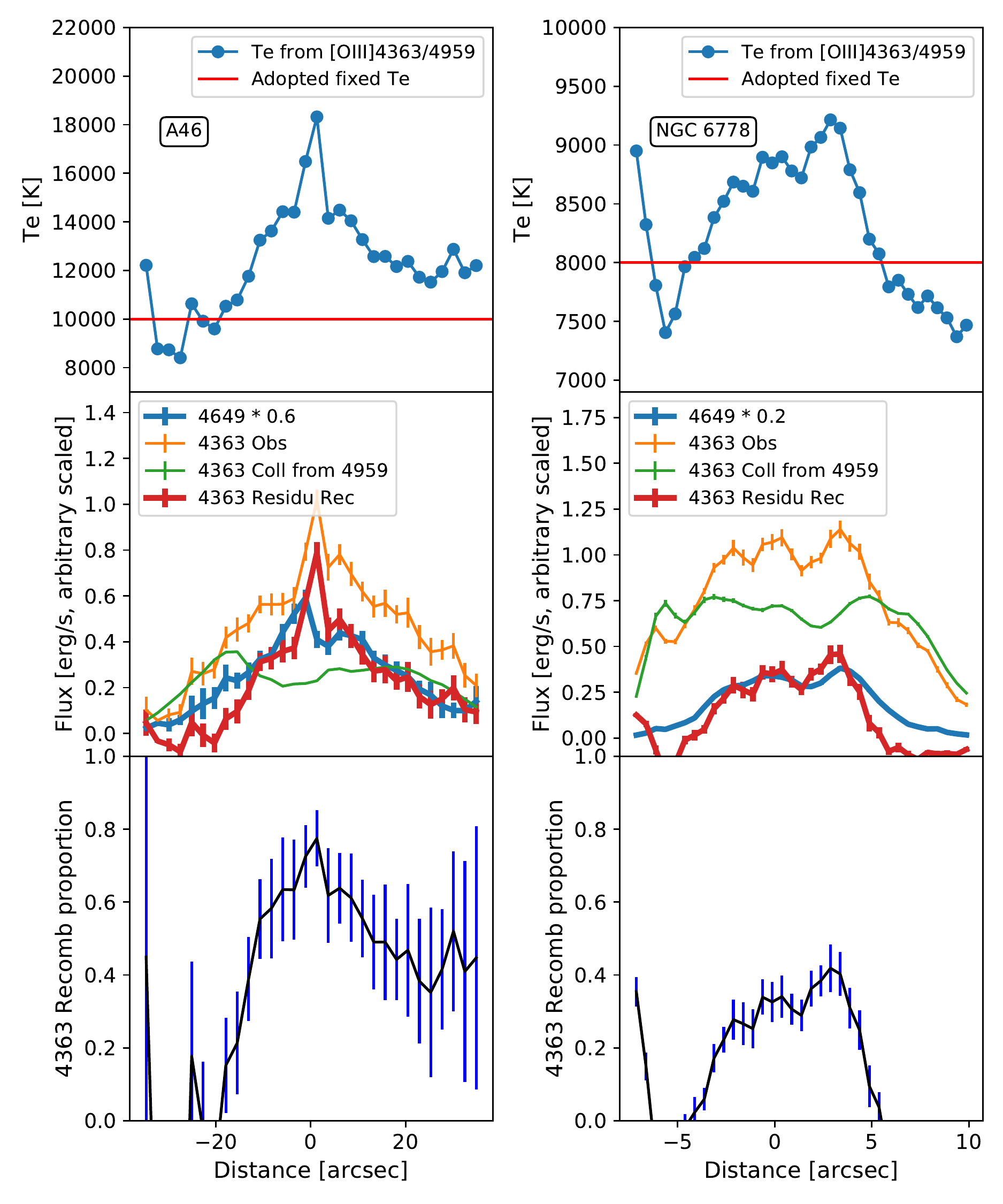}
\caption{Top panels: $T_e$ estimated from [O~{\sc iii}] $\lambda$4363/$\lambda$4959 ratio. The red line represents the adopted $T_e$. Middle panels: Spatial distribution of observed [O~{\sc iii}] $\lambda$4363 (orange), the expected profile of [O~{\sc iii}] $\lambda$4363 emitted by collision assuming the fixed $T_e$ from the upper panels and the profile of [O~{\sc iii}] $\lambda$4959 (green), the residuals from subtracting the expected collisional [O~{\sc iii}] $\lambda$4363 profiles from the observed one (red), and for comparison the O~{\sc ii} $\lambda$4649 profile multiplied by a scale factor (blue). Lower panels: Relative contribution of recombination to the [O~{\sc iii}] $\lambda$4363 line. \label{fig:profiles}}
\end{figure}

Adopting a $n_e$ of 10$^3$~cm$^{-3}$, we estimate (using PyNeb, version 1.1.10) the spatial distribution of $T_e$ in Abell~46 and NGC~6778 from the line ratio [O~{\sc iii}] $\lambda$4363/4959. This is shown in blue in the top left and right panels of Fig.~\ref{fig:profiles}, respectively. In Abell~46, we can see a noticeable increase of the temperature estimation towards the center of the object. In the left and middle panels of Fig.~\ref{fig:siii_lines}, we show MUSE emission line maps of the T$_e$ sensitive [S~{\sc iii}] $\lambda$6312 and [S~{\sc iii}] $\lambda$9068 CELs in NGC\,6778 (Garc\'ia-Rojas et al. in prep.), and in the right panel, we show the $T_e$ map obtained from the ratio of both lines, assuming a constant density $n_e \sim$1,000 cm$^{-3}$. The $T_e$ map shows a roughly constant $T_e$ distribution, with an average $T_e$, weighted by the [S~{\sc iii}] $\lambda$6312 flux of $\sim$8,150 K, implying that S$^{3+}$ recombination emission is not significantly enhanced. This is consistent with previous studies, which suggested that the phenomenon of highly enhanced recombination-line emission is restricted to second-row elements \citep{2003Barlow_209, 2018Wesson_mnra480}.
Following this observational evidence and the bi-metallicity hypothesis, we adopt a constant $T_e$ (red solid line in top panels of Fig.~\ref{fig:profiles}) for the close-to-solar region in Abell~46 and NGC~6778 of 10,000~K and 8,000~K (see above), respectively. The spatial distribution of the observed [O~{\sc iii}]~$\lambda$4363 line, is plotted in orange in the middle panels of Fig. \ref{fig:profiles} for Abell 46 (left) and NGC~6778 (right). To obtain the collisional contribution of [O~{\sc iii}]~$\lambda$4363, we divide the observed emission of the line [O~{\sc iii}]~$\lambda$4959 by the theoretical line ratio [O~{\sc iii}]~$\lambda\lambda$4959/4363 at the adopted constant temperature\footnote{This is obtained with PyNeb, from the ratio of emissivities $\varepsilon$(4959)/$\varepsilon$(4363) using the adopted temperature of 8,000~K and 10,000~K for NGC~6778 and Abell~46 respectively, and a density of 10$^{3}$~cm$^{-3}$. The corresponding ratios are 117 and 51.}. The result is shown in green in the middle panels of Fig.~\ref{fig:profiles}. We then subtract this collisional contribution to the total observed flux of [O~{\sc iii}]~$\lambda$4363 (orange line) to get the possible recombination contribution of $\lambda$4363 from O$^{+3}$ (red line). For comparison, we also show the spatial distribution of O~{\sc ii} $\lambda$4649 ORL (blue line) multiplied by a normalization factor. We can see that the residual spatial profile of [O~{\sc iii}]~$\lambda$4363 (red line) resembles that of the O~{\sc ii} $\lambda$4649 ORL, indicating that the emissivity of the line is dominated by the recombination contribution. In the lower panels of Fig.~\ref{fig:profiles}, we show the spatial distribution of the recombination contribution to the total [O~{\sc iii}]~$\lambda$4363, which reaches up to 70\% and 50\% of the total emission for Abell 46 (left) and NGC~6778 (right), respectively.

The recombination contribution to the [O~{\sc iii}] $\lambda$4363 line may also be estimated using O$^{3+}$ ORLs. Using \citet{1991Pequignot_aap251} via PyNeb, one can for example deduce I([O~{\sc iii}] $\lambda$4363) / I(O~{\sc iii} $\lambda$3762+) increasing from 0.35 to 0.55 (0.45 to 0.7) when $T_e$ increases from 1,000K to 20,000K in case B (case A resp.). \citet{2016Jones_mnra455} report I(O~{\sc iii} $\lambda$3760) = 0.61 (H$\beta$ = 100) for NGC~6778. One can estimate the intensity of the whole V2 multiplet I(O~{\sc iii} $\lambda$3762+) to be $\simeq$ 1.00, leading to a prediction of I([O~{\sc iii}] $\lambda$4363) from recombination to be of the order of 0.4 to 0.5. This is between 20 to 25\% of the observed I([O~{\sc iii}] $\lambda$4363) = 2.07, close to what we obtain for the same PN (see Fig.~\ref{fig:profiles}).


\section{Discussion}
\label{sec:discussion}

In this paper we explored the very crude hypothesis that the $T_e$ of the close-to-solar abundance gas in the central part of the nebula is the same as in the main nebula (the red line showing the adopted value in Fig.~\ref{fig:profiles}). This may not be the case. If one wants to increase the precision in the determination of the [O~{\sc iii}] $\lambda$4363 recombination contribution, one needs to make a detailed photoionization model of the object. This requires a good atmosphere model for the ionizing source in order to correctly reproduce the heating of the nebula. One also may need to take into account the presence of dust and its properties, to accurately compute the balance between the heating and the cooling in the inner part of the nebula. This is totally out of the scope of the simple ``proof of concept" presented in this letter.

One can also question the precision of the atomic data involved in the different parts of the emission calculus, especially the radiative recombination at low temperature.
Although the dielectronic recombination seems to vanish at low temperature (Fig.~\ref{fig:alphas}), it has recently been pointed out that the effect of ``exotic'' atomic processes like {\it Rydberg Enhanced Recombination (RER)} could be very important in these regimes. RER could thus have an impact on the predicted ionization balance and, hence, on the derived ionic abundances \citep[][]{2019Nemer_apjl_887}, changing $n(O^{3+})$ in Eq.
~\ref{eq:I4363b_12}. The residual obtained in Sec.~\ref{sec:biabund} and associated to the $O^{+}$ recombination can also include a contribution from RER, as such an exact computation of the RER-based emission will be important in understanding the entirety of the [O~{\sc iii}] $\lambda$4363 emission.

Regarding observations, it is becoming increasingly clear that for a complete understanding of this problem, a combination of detailed photoionization models with deep IFU observations might improve the situation. In the case of extreme ADF PNe, were two different plasma components coexist, the Balmer and Paschen jumps might not be indicative of any real gas temperature, as they are only a weighted mean of two very different phases of gas. Similarly, the recombination lines (e.g. O~{\sc ii} or H~{\sc i}) are not telling us the value of the ionic abundance ratio O$^{2+}$/H$^+$, as it is impossible to determine what fraction of H$^+$ actually comes from the cold region. The exact weight of the H-poor zone can only be constrained through detailed photoionization models. From the comparison of theoretical models with observations, one can obtain the physical properties ($T_e$, $n_e$, mass, abundances) of the two plasma components that reproduce the observed spectra. However, a detailed treatment of the physics has revealed that the ADF might be only a rough estimate of this discrepancy and is unlikely to provide ``real'' information on the ORL/CEL abundance ratios \citep[see ][]{2020Gomez-Llanos}.


\section*{acknowledgments}

The authors thank the referee for their comments. This paper is based on observations made with ESO Telescopes at the Paranal Observatory under programme IDs 093.D-0038 and 097.D-0241, and on observations made with the William Herschel Telescope operated on the island of La Palma by the Isaac Newton Group of Telescopes in the Spanish Observatorio del Roque de los Muchachos of the Instituto de Astrof\'isica de Canarias. VG-L, CM acknowledges support from projects CONACyT / CB2015 - 254132 and UNAM / PAPIIT - IN101220.
JG-R acknowledges support from an Advanced Fellowship from the Severo Ochoa excellence program (SEV-2015-0548). JG-R, DJ and RC acknowledge support from the State Research Agency (AEI) of the Spanish Ministry of Science, Innovation and Universities (MCIU) and the European Regional Development Fund (FEDER) under grant AYA2017-83383-P.

\section*{Data availability}
The original data used in the paper are available under request to the authors.

\input{output.bbl}

\bsp	
\label{lastpage}
\end{document}